\documentclass[epj]{webofc}
\usepackage[utf8]{inputenc}
\usepackage[varg]{txfonts}   
\usepackage{booktabs}
\usepackage{xcolor}
\usepackage{wrapfig}
\definecolor{darkred}{rgb}{0.4,0.0,0.0}
\definecolor{darkgreen}{rgb}{0.0,0.4,0.0}
\definecolor{darkblue}{rgb}{0.0,0.0,0.4}
\usepackage[bookmarks,linktocpage,colorlinks,
    linkcolor = darkred,
    urlcolor  = darkblue,
    citecolor = darkgreen]{hyperref}
\usepackage{graphicx}
\usepackage{subcaption}
\wocname{EPJ Web of Conferences}
\woctitle{Lattice2017}
\newcommand{\beq}{\begin{equation}}
\newcommand{\eeq}{\end{equation}}
\newcommand{\bea}{\begin{eqnarray}}
\newcommand{\eea}{\end{eqnarray}}
\newcommand{\tr}{\mathrm{tr}}
\newcommand{\mc}{\mathcal}
\newcommand{\la}{\langle}
\newcommand{\ra}{\rangle}
\newcommand{\upa}{\uparrow}
\newcommand{\downa}{\downarrow}
\begin{document}
%
\selectlanguage{english}
\title{%
Particle Projection Using a Complex Langevin Method
}
\author{%
\firstname{Christopher R.} \lastname{Shill}\inst{1} \and
\firstname{Joaqu\'{\i}n E.} \lastname{Drut}\inst{1}
}
\institute{
Department of Physics and Astronomy, University of North Carolina, Chapel Hill, NC, 27599, USA
}
\abstract{
  Using complex stochastic quantization, we implement a particle-number 
  projection technique on the partition function of spin-$1/2$ fermions at finite temperature on the lattice.
  We discuss the method, its application towards obtaining the thermal properties of finite Fermi systems in three
  spatial dimensions, and results for the first five virial coefficients of one-dimensional, attractively interacting
  fermions.
}
\maketitle
\section{Introduction}\label{intro}

Calculating the thermodynamics of any quantum many-body system is a challenging but generally interesting problem for
a variety of applications. By far the main roadblock in Monte Carlo calculations of such systems is the sign problem,
whose impact affects fields ranging from ultracold atoms~\cite{Review1,Review2} to QCD~\cite{CL1}. In the last few years, a concerted effort to tackle this issue has been taking place, as several techniques
have been put forward to that effect (see e.g.~\cite{GattringerSignReview}). In this work, we make use of one of those ideas, namely that of complex 
stochastic quantization (i.e. we use the complex Langevin approach) to explore the feasibility of particle-number projection
in the calculation of the thermal properties of finite clusters and in the numerical determination of virial coefficients.

We begin by recalling that the equilibrium thermodynamics is encoded in the grand canonical partition function,
\bea
\mathcal Z = \tr \left[e^{-\beta (\hat H - \mu \hat N)}\right],
\eea 
where $\hat H$ is the Hamiltonian, $\hat N$ is the particle number operator, $\beta$ is the inverse temperature, and $\mu$ is the chemical potential.
This in itself is nearly impossible to solve for essentially all interacting cases. To gain a better understanding, in particular in dilute regimes, we may expand 
about $z=e^{\beta \mu} = 0$, where $z$ is the fugacity. Such a low-fugacity expansion is what we call the virial expansion. The corresponding coefficients
accompanying the powers of $z$ in the expansion of the grand-canonical partition function $\mathcal{Z}$ are the $N$-particle partition functions;
specifically, for spin-$1/2$ fermions,
\bea
\mathcal Z =  \sum_{n_{\upa},n_{\downa}=0}^{\infty} z_{\upa}^{n_{\upa}} z_{\downa}^{n_{\downa}} Q_{n_{\upa},n_{\downa}}
\eea 
where $z_{\upa} = e^{\beta \mu_{\upa}}$ and $z_{\downa} = e^{\beta \mu_{\downa}}$ are the fugacity for the spin-up and spin-down particles, respectively, and $Q_{n_{\upa},n_{\downa}}$ are the canonical partition functions for $n_{\upa} + n_{\downa}$ particles.

To further analyze the thermodynamics of many-body systems we can study the grand-canonical potential, $- \beta \Omega = \ln \mc Z$. Again, expanding in powers of $z_{\upa}$ and $z_{\downa}$ we achieve an expansion about the dilute limit. This yields,
\bea
-\beta \Omega = \ln {\mathcal Z} = Q_1 \sum_{n_{\upa},n_{\downa}=1}^{\infty} b_{n_{\upa},n_{\downa}} z_\upa^{n_{\upa}} z_\downa^{n_{\downa}}
\eea
Where $b_{n_{\upa},n_{\downa}}$ are the ``virial coefficients'' of order $N = n_{\upa} + n_{\downa}$. These are of particular interest as they give access to the equation of state (EoS), and therefore to observables for these many-body systems (i.e. density, pressure, compressibility, etc.). Moreover, current experiments with
ultracold atoms have gained increasing access to finely controllable parameters such as temperature, coupling, polarization, mass imbalance, and trap shape (harmonic, hard wall, lattices), which allows direct comparison of thermodynamic predictions in the dilute limit, which is the focus of this work.

In this contribution, we show our progress towards accessing $Q_{n_{\upa},n_{\downa}}$ for the universal regime called the unitary Fermi gas~\cite{ZwergerBook}, with the aim of characterizing properties of finite systems, i.e. those derived from the Helmholtz free energy $- \beta F = \ln Q_{n_{\upa},n_{\downa}}$ of finite clusters. Our second main goal is to use a similar approach to compute virial coefficients, $b_{n_{\upa},n_{\downa}}$, at various orders for the one-dimensional Fermi gas with attractive, short-range interactions. To that end, we use a variant of the complex Langevin method as a means of stochastic quantization.

\section{Formalism}\label{form}
\subsection{Preliminaries}

Below, we will make use of the path integral representation of the grand-canonical partition function, namely
\bea
\mathcal Z = \int \mathcal D \sigma \det M[\sigma],
\eea
where $\sigma$ is a Hubbard-Stratonovich field representing a two-body interaction and the system is assumed to be non-relativistic. 
The specific dynamics of the system at hand is encoded in the matrix $M[\sigma]$. In particular for a two-species non-relativistic system, it is not
difficult to show~\cite{QMCReview1} that
\bea
\det M[\sigma] = \det(\mathbb I + z_\uparrow U[\sigma]) \det(\mathbb I + z_\downarrow U[\sigma]),
\eea
where $U[\sigma]$ does not depend on the chemical potential. The above equation is the source of the so-called sign problem in polarized non-relativistic 
matter, as $U[\sigma]$ is real for attractive interactions but $z_\uparrow \neq z_\downarrow$
leads to different determinants for each species (generally of different, fluctuating sign).
 In the sections below, we implement a method to compute canonical partition functions and virial coefficients, both based on the idea
of particle projection, which is in essence a Fourier transform on a variable $\phi$ if the fugacity is replaced by taking $z\to z e^{i\phi}$. Such a transformation, 
which makes $z$ a complex variable, introduces a phase problem in conventional lattice Monte Carlo approaches.

\subsection{Complex Langevin and Stochastic Quantization}\label{CL}

The idea of stochastic quantization is that a random process that obeys the Langevin equation
\bea
\frac{\partial \sigma (\theta)}{\partial \theta} = - \frac{\delta S[\sigma]}{\delta \sigma(\theta)} + \eta,
\eea
will yield configurations $\sigma$ that are distributed according to $e^{-S[\sigma]}$.
Here, $\theta$ is Langevin time and $\eta$ is gaussian random noise (see e.g.~\cite{CL1}).
We assume our Hubbard-Stratonovich field $\sigma$ is real. However, when $S[\sigma]$ is complex, the Langevin equation naturally implies that $\sigma$ must 
be complexified into $\sigma = \sigma^R +i \sigma^I$. Thus, stochastic quantization requires complexified fields and the (spacetime local) Langevin equation now has both real 
and imaginary parts, 
\bea
\sigma^R(n+1) &=& \sigma^R(n) + \epsilon \mathit{K}_{}^R(n)+\sqrt{\epsilon} \eta_{}(n),  \\
\sigma^I(n+1) &=& \sigma^I(n) + \epsilon \mathit{K}_{}^I(n),
\eea
where the Langevin time has been discretized as $\theta = n \epsilon$, and $\epsilon$ is the time step. Again, $\eta$, is the real Gaussian noise such that 
$\la \eta(n) \ra = 0$. The drift terms are now specified as,
\bea
\mathit{K}_{a,x}^R &=& - \mathrm{Re} \left[ \left . \frac{\delta S}{\delta \sigma} \right |_{\sigma \rightarrow \sigma^R + i \sigma^I}  \right], \\
\mathit{K}_{a,x}^I &=& - \mathrm{Im} \left[ \left . \frac{\delta S}{\delta \sigma} \right|_{\sigma \rightarrow \sigma^R + i \sigma^I}  \right].
\eea
%

\subsection{Particle Projection Method}\label{PP}

As a starting point, we use the auxiliary-field representation of the grand-canonical partition function on the lattice,
which reads
\bea
\label{Eq:Zpathintegral}
\mathcal Z &=& \int {\mathcal D} \sigma \det(\mathbb I + z_\uparrow U[\sigma]) \det(\mathbb I + z_\downarrow U[\sigma]),
\eea
where $\sigma$ is the Hubbard-Stratonovich auxiliary field, and $U[\sigma]$ encodes all the properties
of the Hamiltonian of the system. Expanding in terms of the low-fugacity limit we rewrite
\bea
\mathcal Z =  \sum_{n,m}^{\infty} z_{\upa}^{n} z_{\downa}^{m} Q_{n,m}.
\eea 
Particle projection consists in implementing delta functions that select $n_{\upa}$ and $n_{\downa}$ in the above sum, that is $\delta_{n_{\upa},n}$ and $\delta_{n_{\downa},m}$. To that end, we first take $\mathcal{Z}[z_{\upa},z_{\downa}] \rightarrow \mathcal{Z}[z_{\upa} e^{i \phi_{\upa}},z_{\downa} e^{i \phi_{\downa}}]$, such that our fugacity expansion of $\mathcal{Z}$ becomes,
\bea
\mathcal{Z}[z_{\upa} e^{i \phi_{\upa}},z_{\downa} e^{i \phi_{\downa}}]= \sum_{n,m}^{} z_{\upa}^{n} e^{i n \phi_{\upa}} z_{\downa}^{m} e^{i m \phi_{\downa}} Q_{n,m}.
\eea
We may thus project out the desired individual partition functions $Q_{n_{\upa},n_{\downa}}$ using a Fourier integral in the usual way, namely
\bea
&& \!\!\!\!\!\!\!\!\int_{0}^{2 \pi} \frac{d \phi_{\uparrow}}{2 \pi} \frac{d \phi_{\downarrow}}{2 \pi} 
\mathcal{Z}[z_{\upa} e^{i \phi_{\upa}},z_{\downa} e^{i \phi_{\downa}}] z_{\upa}^{-n_{\upa}} e^{-i n_{\upa} \phi_{\upa}} z_{\downa}^{-n_{\downa}}e^{-i n_{\downa} \phi_{\downa}} = \\
&& = \int_{0}^{2 \pi} \frac{d \phi_{\uparrow}}{2 \pi} \frac{d \phi_{\downarrow}}{2 \pi} 
\left( \sum_{n_{\upa},n_{\downa}}^{} 
z_{\upa}^{n} e^{i n \phi_{\upa}} z_{\downa}^{m} e^{i m \phi_{\downa}} Q_{n,m}  \right) 
z_{\upa}^{-n_{\upa}} e^{-i n_{\upa} \phi_{\upa}} z_{\downa}^{-n_{\downa}}e^{-i n_{\downa} \phi_{\downa}} \\
&& = \sum_{n_{\upa},n_{\downa}}^{} z_{\upa}^{n_{\upa} - n} z_{\downa}^{n_{\downa}-m}Q_{n,m} \delta_{n,n_{\upa}} \delta_{m,n_{\downa}} 
=Q_{n_{\upa},n_{\downa}}.
\eea
In practice, we may define $\Phi = (\sigma, \phi_{\upa}, \phi_{\downa})$ as a new auxiliary field variable and perform the Fourier integral as part of stochastic 
process (see below). For this purpose, we include $\phi_{\upa}$ and $\phi_{\downa}$ as part of our action and integral as follows,
\bea
Q_{n,m} = \int \mathcal{D} \Phi \exp \left(- S_{n,m}[\Phi]\right),
\eea
where
\bea
S_{n,m}[\Phi] = - \ln \left[ \det(\mathbb I + z_\uparrow U[\sigma]) \det(\mathbb I + z_\downarrow U[\sigma]) z_{\upa}^{-n} e^{-i n \phi_{\upa}} z_{\downa}^{-m}e^{-i m \phi_{\downa}} \right].
\eea
From this point on, thermodynamic properties of finite clusters can be obtained in the usual way by formally taking derivatives of $Q_{n,m}$ with 
respect to the desired source and computing expectation values in the extended ensemble defined by the variable $\Phi$.

\subsection{Virial Projection}\label{VP}

By definition, the virial expansion of the partition function $\mathcal Z$ can be written as
\bea
\ln \mathcal Z[z] = \mathcal Q_1 \sum_{n=1}^{\infty} b_n z^n,
\eea
where $z$ is the fugacity, $b_n$ are the virial coefficients and $\mathcal Q_1$ is the single-particle partition function.

Based on the above expression, we may use Fourier projection to obtain the virial coefficients:
\bea
b_n = \frac{1}{\mathcal Q_1}\int_0^{2\pi}\frac{d\phi}{2\pi} e^{i \phi n}\ln \mathcal Z [z \to e^{-i\phi}].
\eea
More generally, we may define the function
\bea
b_n(\xi) = \frac{1}{\mathcal Q_1}\int_0^{2\pi}\frac{d\phi}{2\pi} e^{i \phi n}\ln \mathcal Z [z \to \xi e^{-i\phi}] = b_n \xi^n,
\eea
where the last identity is easy to see from our first definition above.

%
This last equation, however, does not help in computing any of the above: while we have a path-integral form for $\mathcal Z$, the presence of the logarithm
makes a direct evaluation impractical.
Nevertheless, inserting the path-integral form of the partition function and differentiating with respect to $\xi$ we obtain a more useful expression: 
\bea
\frac{\partial b_n(\xi)}{\partial \xi} = 
\frac{1}{\mathcal Q_1}\int_0^{2\pi}\frac{d\phi}{2\pi} e^{i \phi n}
\int \mathcal D \sigma \;
P[\sigma,\xi e^{-i \phi}] \;
\tr \left[M^{-1} \partial M/\partial\xi\right],
\eea
where $P[\sigma,z]\equiv {\det M[\sigma,z]  }/{\mathcal Z [z]}$.
Using shorthand notation for the expectation value with respect to the above weight $P$, 
\bea
\label{Eq:EVbn}
\frac{\partial b_n(\xi)}{\partial \xi} = 
\frac{1}{\mathcal Q_1}\int_0^{2\pi}\frac{d\phi}{2\pi} e^{i \phi n}
\left \langle \tr \left[M^{-1} \partial M/\partial\xi\right]\right \rangle_{\phi,\xi} = n b_n \xi^{n-1},
\eea
In practice, we do not need a continuous representation for the Fourier integral, particularly because we cannot include this as part of the stochastic process as in our particle projection method due to the logarithm between these integrals (i.e. the integral over the projecting angles does not appear in the normalization of $P$). Rather, we may use an exact discrete representation via the 
associated angle values $\phi_k = 2\pi k /N_k$, where $k = 0,\dots,N_k-1$ and $N_k$ is the number of discretization points, i.e.
\bea
\label{Eq:EVbnDiscrete}
\frac{\partial b_n(\xi)}{\partial \xi} = 
\frac{1}{\mathcal Q_1}\frac{1}{N_k} \sum_{k=0}^{N_k-1} e^{i \phi_k n}
\left \langle \tr \left[M^{-1} \partial M/\partial\xi\right]\right \rangle_{\phi_k,\xi} = n b_n \xi^{n-1},
\eea

What is interesting about Eq.~(\ref{Eq:EVbn}) is that, by using the complex Langevin method to determine the complex expectation value
as a function of $\phi_k$ and $\xi$, followed by summing over $\phi_k$ with appropriate phases, we may have access to $b_n$ itself. 

In this approach, varying $\xi$ could be very advantageous, as we know the exact $\xi$ dependence of the
final result via the last identity of Eq.~(\ref{Eq:EVbn}). To enhance the sensitivity of the approach, it is preferable to take $\xi$ to be as large as possible,
since the $b_n$ tend to be small numbers. However, we anticipate that taking $\xi \sim 0.1$ or larger could introduce considerable noise in the calculation.
Thus, a priori (and ideally), we expect to find a window of values of $\xi$ makes the approach viable. Our knowledge of the functional dependence 
mentioned above should connect the data points for all $\xi$ with a single curve. In fact, the expected dependence can be accounted for by plotting 
\bea
b_n = \frac{1}{n \xi^{n-1}}\frac{\partial b_n(\xi)}{\partial \xi}
\label{Eq:del_xi}
\eea
as a function of $\xi$ and fitting a constant. If the number of Fourier discretization points $N_k$ is large enough, we expect to
see such a constant behavior beyond the statistical precision. We pursue such an approach below.


\section{Results}\label{Results}

\subsection{Particle Projection Results}
\begin{wrapfigure}{hr}{80mm}
\vspace{-4.0cm}
  \begin{center}
  \includegraphics[scale = 0.35]{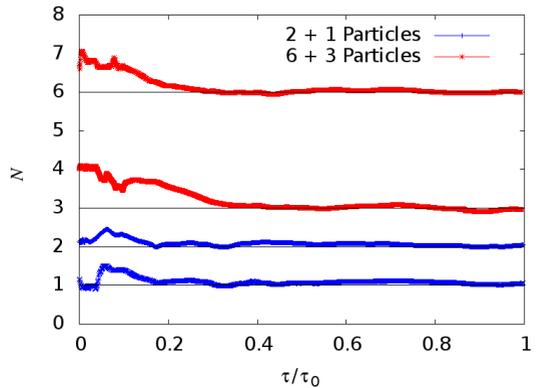}
  \end{center}
  \vspace{-80pt}
  \caption{Running average over Langevin time, $\tau$, where $\tau_0$ is the total number of Langevin time steps, of the total particle numbers, $\la \hat n_{\upa}  \ra$ and $\la \hat n_{\downa}  \ra$. Separate runs are seen to converge to the expected particle number for the $2+1$ case (blue) and the $6+3$ case (red). The data corresponds to $V = N_{x}^{_{}^{}3}$, $N_x = 6$, $\beta = 3.0$, tuned to the unitary limit.}
  \label{fig-1}
\end{wrapfigure}
The simplest and most important verification of this method is computing the projected particle numbers for each flavor. The observable in the
grand canonical ensemble is given by
\bea
\la \hat N  \ra = \la \tr (M^{-1}[\sigma] \partial M [\sigma]/ \partial z)   \ra,
\eea
which is to be averaged over Langevin time. As Langevin time progresses, the running average for the particle numbers of each flavor, $n_{\upa}$ and $n_{\downa}$, 
should converge to the selected values. This is demonstrated in Figure.~(\ref{fig-1}), where we indeed verify that the method converges to the correct solutions.
Figure.~(\ref{fig-1}) is computed in a relatively small spatial volume, $V = 6^{3}$, but at $\beta = 3.0$, which implies $N_\tau = 60$ lattice sites in the time direction. 
Computations for larger volumes (i.e. $N_x = 8, 10 ,12$) are needed to limit finite-size effects. Likewise, larger values of $\beta$ are needed to extrapolate to the
continuum limit (see below).

\subsection{Virial Projection Results}

\begin{figure}[h!]
\centering
\begin{subfigure}[t]{0.49 \textwidth}
  \centering
  \includegraphics[width=1.0\linewidth]{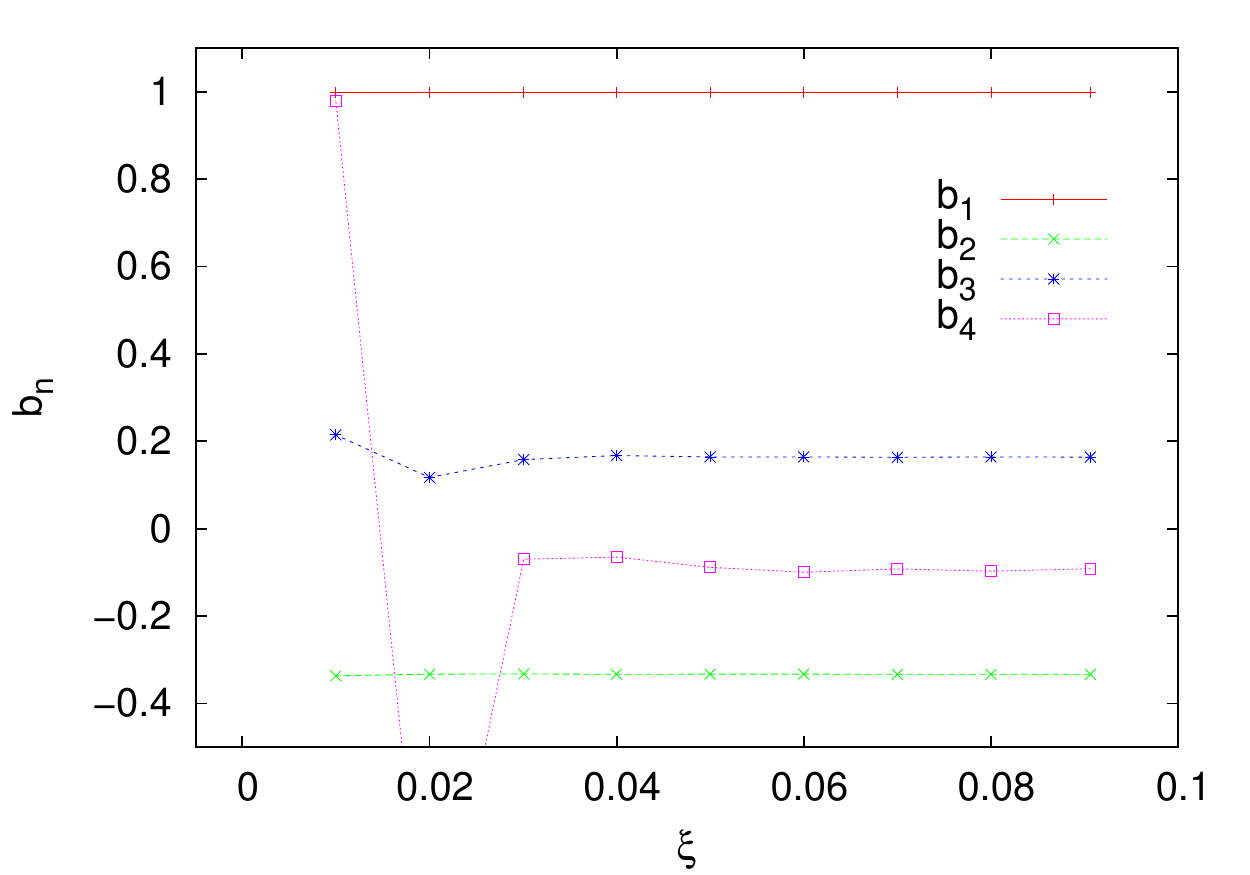}
\end{subfigure}%
\hspace{5pt}\begin{subfigure}[t]{.49\textwidth}
  \centering
  \includegraphics[width=1.0\linewidth]{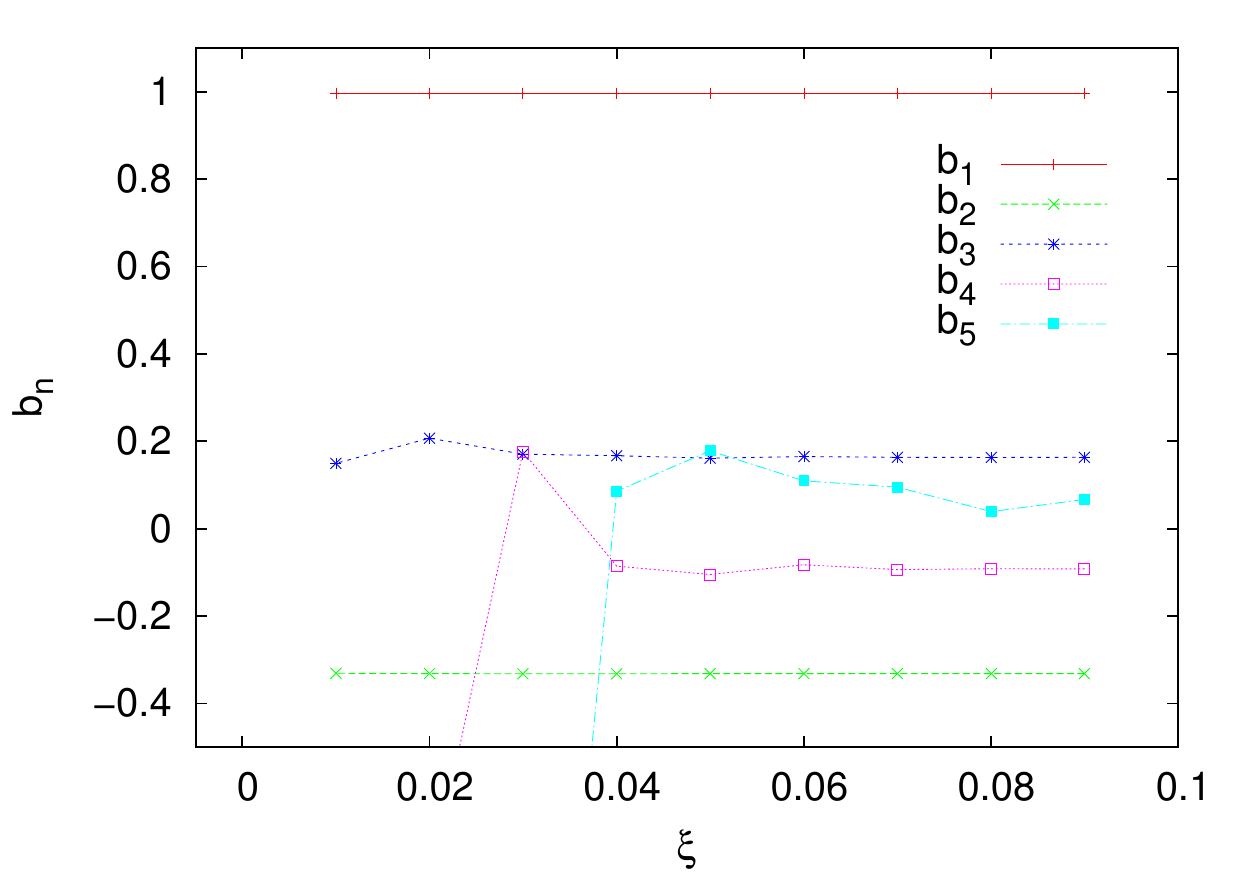}
\end{subfigure}
\caption{Virial coefficients $b_n$, for $n=1,\dots, 5$, as a function of $\xi$ [see Eq.~(\ref{Eq:del_xi})], for $N_x = 31$, $\beta = 8.0$ (left) and $10.0$ (right), and $\lambda = 0.1$.}
\label{fig:bn_proj}
\end{figure}

As mentioned above, the goal of this method is to access the virial coefficients $b_n$ by carrying out calculations for varying values of $\xi$. A first step toward extracting $b_n$ is
to show that the results over the range of $\xi$ yield a constant value, at least in some window. Fig.~\ref{fig:bn_proj}, shows results for $b_n(\xi)$ at $\beta = 8.0$ and $10.0$, both at $\sqrt{\beta} g = 0.1$. We find consistent results for up to order $n=5$ for $\xi = 0.05 - 0.1$, but also find statistical noise at smaller $\xi$, which
is particularly evident in the higher order virial coefficients; the latter correspond to progressively higher frequencies in the Fourier projection. 
Nevertheless, the roughly constant behavior allows us to extract estimates of $b_n$ up to $n=5$. Systematic effects in the number of Fourier points 
$N_k$ and the Langevin time $T$ are currently under study. The data shown here corresponds to $N_k = 100$ and $T = 20$. 

\begin{figure}[h!]
\centering
\begin{subfigure}[t]{0.49 \textwidth}
  \centering
  \includegraphics[width=1.0\linewidth]{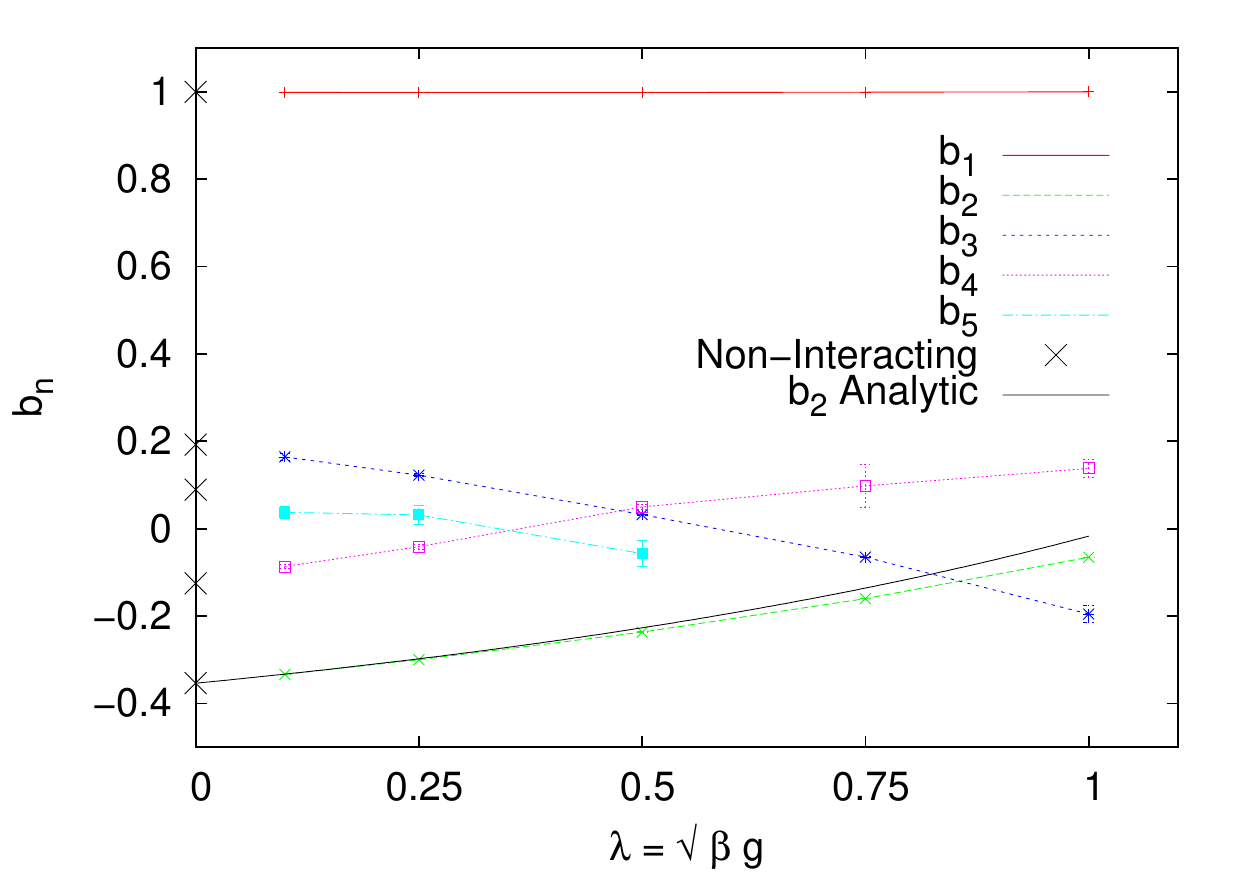}
\end{subfigure}%
\hspace{5pt}\begin{subfigure}[t]{.49\textwidth}
  \centering
  \includegraphics[width=1.0\linewidth]{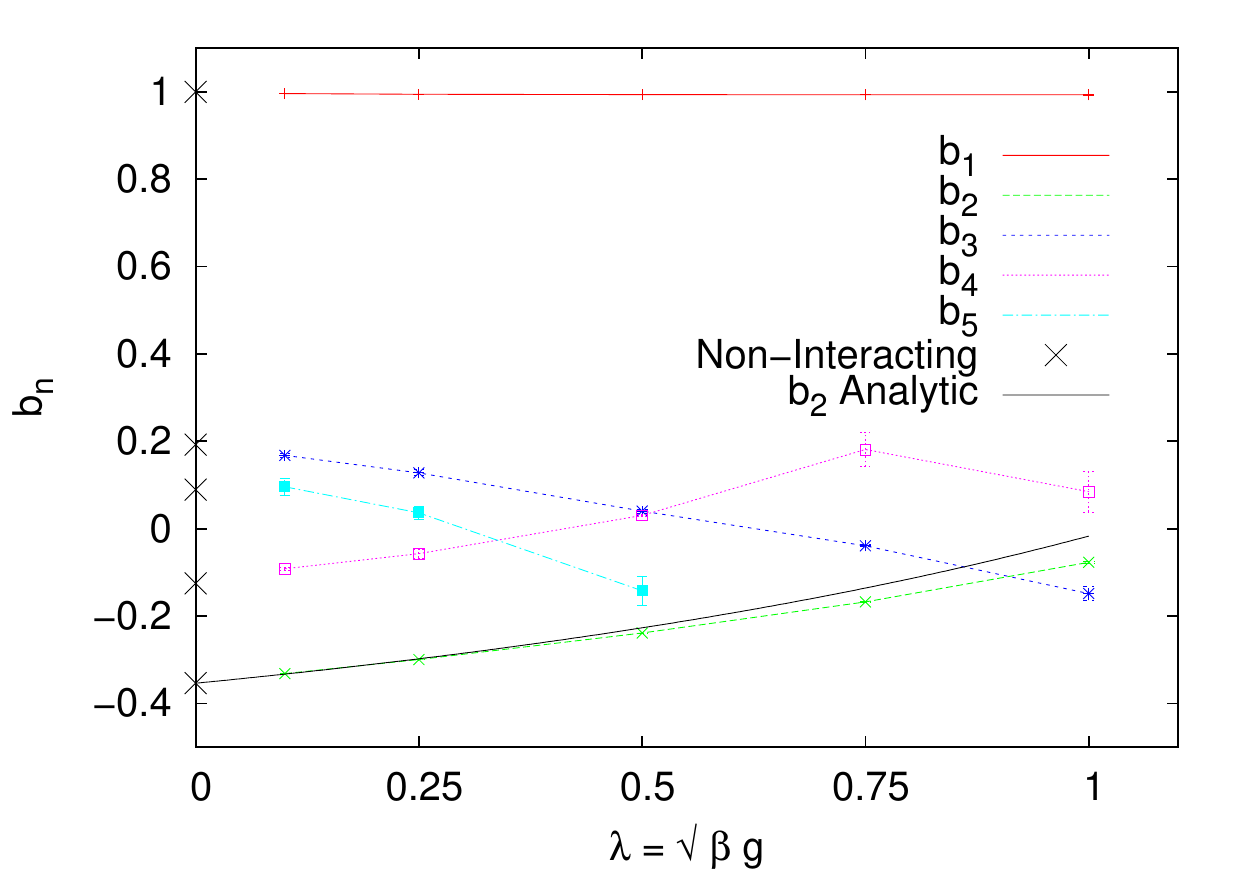}
\end{subfigure}
\caption{Virial coefficients $b_n(\lambda)$ for $n=1,\dots, 5$, at $\beta = 8.0$ (left) and $10.0$ (right), for $N_x = 31$.
The black crosses denote the non-interacting values at the continuum limit. Solid line shows the known continuum-limit result for $b_2$ (see Ref.~\cite{PhysRevA.91.033618}).}
\label{fig:bn_vs_lambda}
\end{figure}

Computing the above results for varying values of the coupling, $\lambda \in [0.1, 1.0]$, and averaging the results over the consistent values (excluding outliers, which are
dominated by noise), we obtain a first estimate of $b_n$ vs. $\lambda$. An extrapolation of our answers to $\lambda = 0$ allows us to compare with the exact solutions for the non-interacting case. Extrapolating in the other direction, i.e. to larger $\lambda$ it may be possible to obtain approximate virial coefficients at strong coupling. These results for 
$b_n$ vs. $\lambda$ at $\beta = 8.0$ and $\beta = 10.0$ are shown in Fig.~\ref{fig:bn_vs_lambda}, together with the exact solution for $b_2(\lambda)$ (see Ref.~\cite{PhysRevA.91.033618}).

Notice in Fig.~(\ref{fig:bn_vs_lambda}) at low coupling our results match extremely well the analytical solution for $b_2(\lambda)$, but that agreement deteriorates 
as $\lambda$ increases. More statistics may be needed at larger coupling to obtain more consistent results.
Fig.~(\ref{fig:bn_vs_lambda}) shows continuous results for most orders of $b_n$, but again with inconsistencies at higher order as $\lambda$ increases. 
Both are encouraging results that point to the possibility of computing high-order virial coefficients and extrapolating to larger coupling with fair accuracy.
This approach, in that sense, complements the conventional, high-precision way to obtain virial coefficients, based on solving the $n$-body problem
and computing canonical partition functions.


\section{Summary and conclusions}

In this contribution we have presented two calculations that use the complex Langevin approach to extract information about 
finite systems from finite-temperature, grand-canonical calculations. We focused on particle-number projection to obtain thermodynamics
via canonical partition functions, and to obtain virial coefficients. We have shown that both methods converge to reasonable results within 
some constraints on various parameters.

More data are needed in order to estimate the free energy for various systems and particle numbers, and to further understand systematic effects. 
To extrapolate to the continuum limit, we must take $1 \ll \lambda_T \ll  N_x$ (where $\lambda_T=\sqrt{2\pi\beta}$ is the thermal wavelength), which amplifies the computation time for 
these calculations as both $\beta$ and $N_x$ must be taken to be large. Due to this limitation, we are exploring ways to optimize these calculations. 
We have shown, however, that particle projection using complex Langevin does converge to the correct state, and with the correct number of particles 
for each flavor.

Furthermore, our method for virial projection has yielded promising results for one-dimensional systems, demonstrating possibilities for 
computing higher order coefficients (i.e. $b_3$, $b_4$, and above) as well as extrapolation to stronger values of the coupling. We are currently exploring finite-size 
effects of this method by varying the inverse temperature, $\beta$, and the volume. In addition, we are investigating the effects of varying the total number of 
Fourier points $N_k$, as decreasing this number likewise decreases the total run time of the simulation (linearly). Increased statistics (longer total Langevin 
time) is needed for improved accuracy for larger virial coefficients, and particularly for larger values of the coupling.

Both particle projection methods have produced promising results that, with more in-depth study, may provide access to
various observables in finite systems that remain unknown and may be directly compared with ultracold atom experiments. 
Looking beyond the latter, the ultimate goal is to perform full-fledged, finite-temperature calculations of finite nuclei with realistic interactions, 
for which the present work provides a proof of principle.


This material is based upon work supported by the
National Science Foundation under Grant No.
PHY{1452635} (Computational Physics Program).


\clearpage
\bibliography{lattice2017}

\begin{thebibliography}{7}

\bibitem{Review1}
S.~Giorgini, L.~Pitaevskii, S.~Stringari, Rev. Mod. Phys. \textbf{80}, 1215
  (2008)

\bibitem{Review2}
I.~Bloch, J.~Dalibard, W.~Zwerger, Rev. Mod. Phys. \textbf{80}, 885 (2008)

\bibitem{CL1}
G.~Aarts, J. Phys. Conf. Ser. \textbf{706}, 022004 (2016)

\bibitem{GattringerSignReview}
C.~Gattringer, K.~Langfeld, International Journal of Modern Physics A
  \textbf{31}, 1643007 (2016)

\bibitem{ZwergerBook}
W.~Zwerger, \emph{BCS-BEC crossover and the Unitary Fermi Gas}
  (Springer-Verlag, Berlin, 2011)

\bibitem{QMCReview1}
J.E. Drut, A.N. Nicholson, J.Phys. \textbf{G40}, 043101 (2013),
  \texttt{1208.6556}

\bibitem{PhysRevA.91.033618}
M.D. Hoffman, P.D. Javernick, A.C. Loheac, W.J. Porter, E.R. Anderson, J.E.
  Drut, Phys. Rev. A \textbf{91}, 033618 (2015)

\end{thebibliography}

\end{document}